\documentclass[11pt,a4paper]{JHEP3}
\usepackage{amsmath}

\title{Back Reaction from Trace Anomaly in RN-blackholes Evaporation\thanks{published in JHEP12(2003)055} }

\author{F. Canfora and G. Vilasi\thanks{Corresponding author: Gaetano Vilasi,
Universit\`{a} di Salerno, Dipartimento di Fisica
''E.R.Caianiello'', Via S.Allende, 84081 Baronissi (Salerno)
Italy; e-mail: vilasi@sa.infn.it, Phone: +39-089-965317, Fax:
+39-089-965275.}\\
Istituto Nazionale di Fisica Nucleare, GC di Salerno, Italy. \\
Dipartimento di Fisica ''E.R.Caianiello'', Universit\`{a} di
Salerno.}

\preprint{ }

\abstract{A model is proposed to describe a transition from a
charged black hole of mass $M$ and charge $Q$ to one of mass
$\overline{M}$ and charge $\overline{Q}$. The basic equations are
derived from the non-vacuum Einstein field equations sourced by
the Coulomb field and by a null scalar field with a nonvanishing
trace anomaly. It is shown that the nonvanishing trace of the
energy-momentum tensor prevents the formation of a naked
singularity.}


\begin{document}


%

One of the most outstanding results of Quantum Field Theory (QFT)
on curved space-time is the Hawking radiation discovery
\cite{Ha74}which strengthened the analogy between blackhole
mechanics and thermodynamics \cite{Ba73}. Unfortunately, since a
complete evaporation would violate unitarity, the consequences of
the Hawking radiation seem to undermine the physical basis either
of QFT or of the general relativity itself. Indeed, the Hawking
particles, and, more generally, quantum sources violating the
energy conditions, are able to violate the third law of blackhole
thermodynamics, eventually giving rise to naked singularities.
Recently, the role of the four dimensional trace anomaly in the
blackhole evaporation has been investigated \cite{CV303}. Here, we
want to analyze the effects of Hawking particles and of trace
anomaly on another important feature of blackhole thermodynamics:
the third law.

The third law, analogue of the third law of thermodynamics, states
that, if the energy-momentum tensor of matter sources satisfies
the \textit{weak energy condition}, then the surface gravity of
the blackhole, that plays the role of the temperature, cannot be
lowered down to zero in a finite series of steps \cite{Is86}. This
is not the case for energy-momentum tensors arising as vacuum
expectation values of quantum fields, as in the Hawking radiation.
Therefore, the third law can be violated during the evaporation
and, more generally, when energy-momentum tensors not satisfying
positivity conditions come into play. This could have important
consequences also for another of main unsolved problems in general
relativity: the \textit{cosmic censorship hypothesis}
(\textit{CCH}) which, roughly speaking, states that singularities
are always hidden behind a horizon. Indeed, if a suitable quantum
field is able to lower the surface gravity of a blackhole down to
zero in a finite time, we could end up with a naked singularity.
This could be the case during the evaporation of a charged
blackhole if one suppose that Hawking particles carry away mass
but not charge to infinity. This supposition is not unlike since
the Hawking particles are massless (the emission of massive
particles is highly suppressed) while charged massless particles
do not exist. Despite many attempts \cite{Jo02} \textit{CCH} is
still unproved. Furthermore, some beautiful exact results in the
spherically symmetric cases \cite{Chri94}, \cite{Da02b} show that
it is possible to construct naked singularity even with reasonable
matter sources, this being an indication of the fact that
\textit{CCH} could need quite stronger hypothesis than the third
law to be proved in general. Some insights into the physics of
\textit{CCH} came from the analysis of particular cases in which
explicit calculations can be carried on. In the interesting cases
of charged blackholes, without or with cosmological constant
\cite{Cha82}, \cite{Bra98}, the stability analysis showed that
naked singularities are unstable against linear perturbations.

Thus, it is quite interesting to study the effects of quantum sources, such as
the Hawking radiation, on the surface gravity of an initially ''dressed''
Reissner-Nordstrom blackhole. Here we propose a toy model to investigate the
following question: \textit{is it possible to transform an initially non
extremal charged blackhole into an extremal one by means of Hawking particles?}

Indeed, the most promising candidates to lower the surface gravity
are energy-momentum tensors arising as vacuum expectation values
of quantum fields since such tensors are able to violate all the
energy conditions\footnote{Because of the Israel theorem
\cite{Is86}, these violations of the energy conditions are
necessary in order to violate the third law.}. Furthermore, they
are not ''strange'' enough to be neglected, being quite generic in
Quantum Field Theory (QFT). We will prove, \textit{via} the exact
Einstein equations, that, in this model, an energy-momentum tensor
describing a null quantum fluid (that is, a null fluid violating
all the energy conditions) with a nonvanishing trace anomaly
cannot lower the surface gravity of the charged blackhole, so that
the evaporation stops. Thus, the trace anomaly, that is a generic
feature of QFT, could save the third law and prevent the formation
of a naked singularity even if the energy-momentum tensor of the
Hawking particles violates all the energy conditions.

\section{The model}

It is often claimed that blackholes evaporation can be explained
only by a full quantum theory of gravity such as the
\textit{superstring} \textit{theory} \cite{Da02} or the
\textit{loop} \textit{quantum} \textit{gravity } \cite{Ti01}.
Recently, more ''conservative'' approaches to the topic, such as
the \textit{effective} \textit{action} \cite{MWZ94,BGVZ94} and the
\textit{renormalization} \textit{group} \textit{method}
\cite{Bo00}, also have given new insights into the physics of the
evaporation. However, until the final stages, the evaporation is
not a strong gravitational field phenomenon because the blackhole
mass decreases slowly with time \cite{Ba81}. Accordingly, the
standard Einstein equations with a suitable matter source can
describe how the metric evolves during the evaporation. Moreover,
dynamical implementations of loop quantum gravity show that
discrete eigenvalues of geometric operators converge in a few
Planck lengths to their semiclassical approximations \cite{Bo01l}
\cite{Bo01}.

Thus, far away (with respect to the Planck length) from the horizon a
semiclassical description should describe carefully the evaporation process.
Therefore, this is an accurate tool for our purposes, since, to study the
evaporation, one has to look at the mass of the blackhole brought at infinity
by Hawking particles. Furthermore, it would be nice to solve the problems
raised by the blackhole evaporation, such as the loss of unitarity, by only
using general relativity and QFT.

The standard \textit{semiclassical program} \cite{Ba81}, \cite{Pa82},
\cite{Yo85}, \cite{Ka94}, \cite{Pa94}, \cite{Ma95}, \cite{Ma00} to approach
the analysis of the back-reaction of Hawking particles on the metric consists,
first, in computing an effective vacuum stress-energy tensor of a null quantum
field on the black hole space-time of interest \cite{Pa82} and, then, in using
this tensor as a source for the Einstein equations \cite{Yo85}, \cite{Ka94}.
In this framework, the results can be easily interpreted in terms of the
unperturbed quantity, thus making manifest the changes of the blackhole
space-time due to Hawking particles \cite{Yo85}, \cite{Ka94}.

However, it is quite hard to obtain the expression of the vacuum stress-energy
tensor of quantum fields on a curved space-time. Usually, in four-dimensional
geometries, it is possible to get approximated expressions for the
energy-momentum tensor only in static situations \cite{Pa82}, while, in
dynamical models, such as the space-time of an evaporating blackhole, it is
usually computed in effective two-dimensional geometries with the refined
techniques available in \textit{string} and \textit{conformal field theory}
\cite{Di92}, \cite{Ja93}, \cite{Pa94}, \cite{No00}, \cite{GKV02} and in
three-dimensional models with the powerful methods provided by
three-dimensional gravity\footnote{Some attempts to take into account the
four-dimensional trace anomaly, in the framework of \textit{dilatonic
gravity,} can be found in \cite{No00}.} \cite{Wi88}, \cite{Wi89}, \cite{Ma89}.
Therefore, in such effective models, it is not possible to take into account
the nontrivial effects of the four-dimensional trace anomaly.

Thus, a toy model is proposed in which it is possible to keep the form of the
stress-energy tensor of quantum fields as general as possible and, at the same
time, to fully take into account the trace of the stress-energy tensor of
Hawking particles. It will be shown, \textit{via} the exact Einstein
equations, that this has important physical consequences.

\section{The Einstein tensor\ \ }

A transition from a Reissner-Nordstrom blackhole of initial mass $M$ and
charge $Q$\ to another one of final mass $\overline{M}$ and charge
$\overline{Q}$\ will be described outside the collapsed matter by means of a
well suited source that carries away mass from the blackhole to infinity.\ The
metric, valid only for $r>R^{\ast}>0$, $R^{\ast}$ denoting a radius larger
than the initial radius of the outer horizon (and much larger than the Planck
scale, as we will explain in more detail later on), will be written as in
\cite{Ba81}:%

\begin{equation}
g=\left(  1-\frac{m(r,v)}{r}\right)  e^{2j(r,v)}dv^{2}-2e^{j(r,v)}%
drdv-r^{2}d\Omega\hbox{.} \label{m1}%
\end{equation}
For $m=2M-Q^{2}/r$ and $j=const$, metric (\ref{m1}) reduces to the
Reissner-Nordstrom metric in Eddington-Finkelstein coordinates.

Since we want to probe the third law in the worst conditions, we will search
for a solution of Einstein equations satisfying the following boundary
conditions:
\begin{align*}
m(r,0)  &  =2M-\frac{Q^{2}}{r},\quad M>\left|  Q\right|  ;\quad m\left(
r,T\right)  =2\overline{M}-\frac{\overline{Q}^{2}}{r};\quad\\
Q  &  =\overline{Q};\quad\overline{M}\geq0;\quad j\left(  r,0\right)
=k_{1};\quad j\left(  r,T\right)  =k_{2},
\end{align*}
where $k_{1}$ and $k_{2}$\ are two constants and at $v=T$ \ the blackhole
reaches its final state. In fact, by taking the final charge equal to the
initial one $Q=\overline{Q}$, we compel the blackhole to loose its mass but
not its charge so, in this way, by choosing a suitable energy-momentum tensor
violating the \emph{weak energy condition}, we can go as near as possible to
the violation of the third law and, eventually, of the \emph{CCH}.

The above metric has the following scalar curvature
\begin{align*}
R  &  =\left\{  -\exp(-j)2r^{2}\partial_{rv}^{2}j+3r\partial_{r}j\partial
_{r}m+m\partial_{r}j\right. \\
&  +2\partial_{r}m-4r\partial_{r}j-2r^{2}\left(  1-\frac{m}{r}\right)
\partial_{r}^{2}j\\
&  \left.  -2r^{2}\left(  1-\frac{m}{r}\right)  \left(  \partial_{r}j\right)
^{2}+r\partial_{r}^{2}m\right\}  \frac{1}{r^{2}}%
\end{align*}
and the nonvanishing independent components of the Einstein tensor are
\begin{align*}
G_{vv}  &  =\frac{\exp\left(  2j\right)  }{r^{3}}\left(  m\partial_{r}%
m-r\exp(-j)\partial_{v}m-r\partial_{r}m\right)  ,\\
G_{vr}  &  =\frac{\exp(j)\partial_{r}m}{r^{2}},\quad G_{rr}=-2\frac
{\partial_{r}j}{r},\\
G_{\theta\theta}  &  =\frac{1}{2}\left\{  -\left(  2r+m\right)  \partial
_{r}j-\exp(-j)2r^{2}\partial_{rv}^{2}j\right. \\
&  +3r\partial_{r}j\partial_{r}m-2r^{2}\left(  1-\frac{m}{r}\right)
\partial_{r}^{2}j\\
&  \left.  -2r^{2}\left(  1-\frac{m}{r}\right)  \left(  \partial_{r}j\right)
^{2}+r\partial_{r}^{2}m\right\}  ,\mathtt{\ }\\
G_{\varphi\varphi}  &  =\sin^{2}\theta G_{\theta\theta}\mathtt{.}%
\end{align*}

\section{The source \ \ }

During an evaporation process, the black hole emits black-body
radiation at the Hawking temperature. Since the emission of
massive particles by the blackhole is strongly suppressed, the
natural choice is to take as source, beyond the electromagnetic
(${\it e.m.}$) energy-momentum tensor $T_{\mu\nu}^{C}$ , a null\
quantum fluid, so that

\begin{equation}
T_{\mu\nu}=g_{\mu\nu}V\left(  v,r\right)  +\rho\left(  v,r\right)  u_{\mu
}u_{\nu}+T_{\mu\nu}^{C}, \label{t1}%
\end{equation}
with $g\left(  U,U\right)  =0$, being $U$ the vector field
naturally associated with the differential $1$-form
$u_{\mu}dx^{\mu}\,\ \ $($u_{\mu }=g_{\mu\nu}U^{\nu}$) and being
$V$  the trace anomaly of quantum fields representing the Hawking
particles.

\ Thus, $T_{\mu\nu}$ has to be interpreted as the sum of the
energy-momentum tensor of the ${\it e.m.}$ field
\[
T_{\mu\nu}^{C}=g_{\mu\nu}\epsilon\frac{Q^{2}}{r^{4}}\mathtt{,}%
\]
with $\epsilon=1$ for $\mu,\nu=1,2$; $\epsilon=-1$ for
$\mu,\nu=3,4 $, and a suitable average over quantum fields that
contribute to the Hawking radiation and bring mass from the
blackhole to infinity. However, we will be interested only in the
asymptotic region ($r\rightarrow \infty$) where Hawking particles,
contributing to the black hole mass decrease, have to arrive.
Thus, Coulomb terms which are of order $1/r^4$ and then do not
contribute to the Bondi mass-whose change we are interested in-
can be neglected. We will take a neutral fluid as source because
in such a way $T_{\mu\nu}$, bringing at infinity mass but not
charge, allows in principle arbitrary violations of either the
third law or the \emph{CCH}.

It is fair to say that as the Hawking temperature increases, the
probability of charged particles production, specially electrons,
increases as well. The inclusion of the corresponding black hole
charge decrease is interesting but rather cumbersome because it is
necessary to take into account the charge density continuity
equation. In Section 6, it will be shown that in a Q-constant
process the trace anomaly is able to prevent the formation of a
naked singularity. In its turn, the formation of a naked
singularity is more difficult in a process in which the black hole
charge decreases. This can be understood by comparing the distance
($\propto\sqrt{M^{2}-Q^{2}}$ ) between the inner and the outer
horizons in the two processes in which, without loss of
generality, the mass variation may be assumed\footnote{This
assumption is lawful because the main equations, (\ref{F1}) and
(\ref{F2}), of the model do not depend on the explicit form of the
energy density $\rho$ of Hawking particles. Therefore, the
function $\rho$ can assume arbitrarily large and negative values
and can include the contribution to the black hole mass decrease
of all possible kinds of particles participating in the
evaporation} to be the same. In absence of numerical computations,
we are still unable to predict the percentage of mass lost in a
process in which the charge Q is allowed to decrease. At the
moment, we only know that naked singularities do not occur and
that all results also hold in the limit $Q\mapsto 0$, that is the
trace anomaly stops the evaporation of a Schwarzschild black hole.

Since no constraints will be imposed on $\rho$ and $V$, the tensor
$T_{\mu\nu}$ needs not to satisfy neither the \emph{weak} nor the
\emph{null} energy conditions, thus providing us the violations we
are searching for. To see this, let us write the \emph{weak energy
condition}:
\[
T_{\mu\nu}n^{\mu}n^{\nu}\geq0
\]
for every non space-like \emph{future directed} vector field $n^{\nu}$.

In the case of the energy-momentum tensor in Eq. (\ref{t1}) the above relation
reads:
\[
g\left(  n,n\right)  V\left(  v,r\right)  +\left[  g\left(  U,n\right)
\right]  ^{2}\rho\left(  v,r\right)  \geq0,
\]
then we see that $T_{\mu\nu}$\ obeys or not the \emph{weak energy condition}
according to the explicit expressions of $\rho$ and $V$. We can say that the
$T_{\mu\nu}$ in Eq. (\ref{t1}) represents a truly \emph{quantum} source in two
respects: firstly, it encodes the effects of the trace anomaly; secondly, it
do not satisfy in general the \emph{weak energy condition}. However, as we
will show later on, our results, not depending on the explicit form of $\rho$
and $V$, hold irrespectively of the energy conditions fulfilled by $T_{\mu\nu
}$.

The \emph{null energy condition} is:
\[
T_{\mu\nu}l^{\mu}l^{\nu}\geq0
\]
for every light-like future directed vector field $l^{\nu}$, and, in our
model, it reads:
\[
\left[  g\left(  l,U\right)  \right]  ^{2}\rho\left(  v,r\right)  \geq0,
\]
so that, when $\rho\geq0$, the \emph{weak} \emph{energy condition} can still
be violated but the \emph{null} cannot. Furthermore, it is easy to see that,
when $\rho\geq0$, $T_{\mu\nu}$ can be thought as the energy-momentum tensor of
a null scalar field with trace anomaly:
\begin{equation}
T_{\mu\nu}=\partial_{\mu}\varphi\partial_{\nu}\varphi+g_{\mu\nu}V\left(
\varphi\right)  \mathtt{.} \label{t2}%
\end{equation}
First of all, let us notice that the only properties of $T_{\mu\nu}$ (see
below) that we will need are:
\[
T_{rv}^{2}-T_{rr}T_{vv}=\left(  g_{rv}\right)  ^{2}V^{2},\quad g^{\mu}{}^{\nu
}T_{\mu\nu}=4V,
\]
and either (\ref{t1}) or (\ref{t2}) satisfy these relations. Then,
when $\rho\geq0$, the $r-r$ components of both energy momentum
tensors (\ref{t1}) and (\ref{t2}) are greater than zero:
\begin{align*}
\rho\left(  u_{r}\right)  ^{2}  &  \geq0\\
\left(  \partial_{r}\varphi\right)  ^{2}  &  \geq0\mathtt{.}%
\end{align*}
Thus, as far as we are concerned, when $\rho\geq0$, there is no difference in
treating with the $T_{\mu\nu}$ given by Eq. (\ref{t1}) or the one in Eq.
(\ref{t2}).

\section{The Einstein equations \ \ }

Up to non radiative terms, that do not contribute to the
asymptotic Hawking flux, the Einstein equations $G_{\mu\nu}=\kappa
T_{\mu\nu}$\ give:
\begin{align}
\kappa^{-1}G_{vv}  &  =V\left(  1-\frac{m}{r}\right)  \exp\left(  2j\right)
+\rho\left(  u_{v}\right)  ^{2},\nonumber\\
\kappa^{-1}G_{vr}  &  =-\exp\left(  j\right)  V+\rho u_{r}u_{v},\mathtt{\ }%
\ \kappa^{-1}G_{rr}=\rho\left(  u_{r}\right)  ^{2},\label{stop}\\
\kappa^{-1}G_{\theta\theta}  &  =g_{\theta\theta}V,\mathtt{\ \ }-R=4\kappa
V.\nonumber
\end{align}

The last two equations allow us to write the scalar curvature as:
\[
R=\frac{4}{r^{2}}\left[  \left(  m-r\right)  \partial_{r}j+\partial
_{r}m\right]  .
\]
Moreover, thanks to the algebraic properties of $T_{\mu\nu}$ and taking into
account that $g\left(  U,U\right)  =0$, we easily get
\begin{equation}
T_{vr}^{2}-T_{vv}T_{rr}=\left(  g_{vr}\right)  ^{2}V^{2}\mathtt{{.} }
\label{preal}%
\end{equation}
The above relation is a consequence of the non vanishing trace anomaly and it
allows to find a closed system of equations for $m$ and $j$ that does not
depend on the explicit form of $\rho$ and $V$. Indeed, by the field equations,
Eq. (\ref{preal}) implies the following constraint on the Einstein tensor:
\begin{equation}
G_{vr}^{2}-G_{vv}G_{rr}=\kappa^{2}\left(  g_{vr}\right)  ^{2}V^{2}=\left(
g_{vr}\right)  ^{2}\left(  \frac{R}{4}\right)  ^{2} \label{algebra}%
\end{equation}
which gives
\begin{equation}
\left(  m-r\right)  ^{2}\left(  \partial_{r}j\right)  ^{2}=-2r\exp\left(
-j\right)  \partial_{r}j\partial_{v}m \label{separa}%
\end{equation}
and this relation depends only on the fact that $V\neq0$, no matter how $\rho$
and $V$ really look like. Then, $\rho$ and $V$ can be positive, negative, and,
moreover, can be either local or nonlocal functions, thus representing a truly
quantum source, the main equations of the model not depending on their
detailed form.

Let us firstly analyse the case $\partial_{r}j=0$. To begin with, thanks to
the Einstein equation $G_{rr}=\kappa T_{rr}$, we get
\[
\ \partial_{r}j=0\Rightarrow G_{rr}=0\Rightarrow T_{rr}=0\Rightarrow
u_{r}=0\Rightarrow U^{v}=0,
\]
since it is well known that, if $\rho=0$ (that is, of course, the other
solution of $T_{rr}=0$) the unique solution is the Reissner-Nordstrom-De
Sitter one. Then, \emph{via} the Bianchi identities, from $U^{v}=0$ it follows
that:
\[
L_{U}V=0\Rightarrow\partial_{r}V=0
\]
where $L_{U}$ is the Lie derivative along $U$. Thus we see that, in this case,
$V$ depends only on $v$. From the field equations $G_{\theta\theta}%
=g_{\theta\theta}V$ and $G_{vr}=\kappa T_{vr}$\ we get
\[
V=-\frac{\partial_{r}^{2}m}{2r}=-\frac{\partial_{r}m}{r^{2}}\mathtt{.}%
\]
The above relations imply that
\[
m=I_{1}(v)r^{3}+I_{2}(v);\mathtt{\ }\ V=-3I_{1}(v).
\]
However, this expression, unless $I_{1}(v)=0$, gives a diverging Bondi flux
\cite{Bo62} (see below Eq. (\ref{bofu})), so that we have to take $I_{1}(v)=0
$, that is $V=0$, this yielding a vanishing trace anomaly against the
hypothesis $V\neq0$.

Thus, in order to describe the physical effects of a non vanishing trace
anomaly, we have to assume $\partial_{r}j\neq0$. Eventually, by using
$\partial_{r}j\neq0$ in Eq. (\ref{separa}), the equations for $m$ and $j$
reduce to
\begin{equation}
\frac{\exp\left(  j\right)  \partial_{r}j}{2}=\partial_{v}\left(  \frac
{r}{m-r}\right)  , \label{F1}%
\end{equation}%
\begin{align}
3m\partial_{r}j  &  =-\exp\left(  -j\right)  2r^{2}\partial_{rv}^{2}j+\left(
3r\partial_{r}j-2\right)  \partial_{r}m\nonumber\\
&  -r\partial_{r}^{2}m-2r^{2}\left(  1-\frac{m}{r}\right)  \left(
\partial_{r}^{2}j+\left(  \partial_{r}j\right)  ^{2}\right)  , \label{F2}%
\end{align}
where it is worth to note that only Eq.(\ref{F1}) depends on the assumption
$\partial_{r}j\neq0$.

\section{The boundary conditions \ \ }

We want to describe a transition from two charged blackhole (of
the same charge) by the means of the energy-momentum tensor, whose
role is to take away energy from the blackhole to infinity. Thus,
if the space-time is initially asymptotically flat, so it will be
during all the evaporation process. Moreover, the asymptotically
flat region is the most important one since we want to evaluate
the amount of blackhole's mass that the Hawking particles are able
to bring at infinity. For this reason, we are only interested in
the boundary conditions in the region where $r\rightarrow\infty$,
while, with this toy model, we cannot explore the small $r$-region
where quantum-gravitational effects come into play so that we will
only consider the region where $r>R^{\ast}$, $R^{\ast}$ being a
radius much larger than the Planck scale.

Then, let us look at the\ conditions to impose on the coefficients of the
metric (\ref{m1}) that will be assumed to be continuous up to the second
derivatives: $m,j\in C^{2}\left(  \left]  0,T\right[  \times\left]  R^{\ast
},\infty\right[  \right)  $. First of all, the asymptotic flatness implies
that $j$ cannot diverge for $r\rightarrow\infty$, so that $\left|  j\right|
\leq K_{1}<+\infty$. The same is true for $m$. Therefore, either $m$ or $j$
can be expanded there in series of $\frac{1}{r}$:%

\begin{equation}
j\underset{r\rightarrow\infty}{\sim}\sum_{n\geq0}\frac{c_{n}(v)}{r^{n}},\quad
m\underset{r\rightarrow\infty}{\sim}\sum_{n\geq0}\frac{b_{n}(v)}{r^{n}},
\end{equation}
where the coefficients $c_{n}(v)$\ and $b_{n}(v)$ are bounded continuous
functions. Thus, we immediately see that the leading terms of $\partial_{r}j$
and $\partial_{r}m$ in the limit $r\rightarrow+\infty$ (namely, the terms that
do contribute to the Bondi mass) fulfil the bounds
\begin{align}
&  \left|  \partial_{r}j\right|  \underset{r\rightarrow+\infty}{\lesssim}%
K_{2}r^{-2}\label{con1}\\
&  \left|  \partial_{r}m\right|  \underset{r\rightarrow+\infty}{\lesssim}%
K_{2}r^{-2},\label{con2}%
\end{align}
where $K_{2}$ is a positive constant. Furthermore, the Bondi mass \cite{Bo62}
$\Phi_{B}$ of the metric (\ref{m1}) is:
\begin{align}
\Phi_{B}(v) &  =-\frac{1}{8\pi}\underset{r\rightarrow\infty}{\lim}\int_{S_{r}%
}g^{rv}\Gamma_{vv}^{v}dS\nonumber\\
&  =\frac{1}{8\pi}\underset{r\rightarrow\infty}{\lim}\int_{S_{r}}\left[
e^{j}\left[  m+2r\left(  r-m\right)  \partial_{r}j\right]  +2r^{2}\partial
_{v}j-r\partial_{r}m\right]  d\Omega,\label{bofu}%
\end{align}
where $S_{r}$ is the sphere $\left\{  r=const,v=const\right\}  $. Thus, we see
that the previous fall-off conditions ensure a finite Bondi flux provided we
also take
\[
\left|  \partial_{v}j\right|  \underset{r\rightarrow+\infty}{\lesssim}%
K_{3}r^{-2},
\]
where $K_{3}$ is a positive constant that can be chosen to be equal to $K_{2}%
$. Moreover, since the total Bondi flux is bounded from above (otherwise the
energy conservation would be violated), we have to impose that
\begin{equation}
\left|  \int_{0}^{T}\partial_{v}\Phi_{B}(v)dv\right|  <+\infty\label{tobo}%
\end{equation}
from which it follows
\[
\left|  \int_{0}^{T}\left(  \partial_{r}j\right)  dv\right|  \underset
{r\rightarrow+\infty}{\leq}K_{4}r^{-2}\mathtt{,}%
\]
$K_{4}$ being a positive constant. It is worth to note here that the
requirement of a finite total Bondi flux is a necessary condition to ensure
the conservation of energy, since the total Bondi flux is nothing but the
difference between the final and the initial mass of the blackhole. For these
reason, the above relation has to be fulfilled even in the limit
$T\rightarrow+\infty$. Moreover, the condition (\ref{tobo}) imposes a new
constraint only on $\partial_{r}j$ while, as far as the other terms are
concerned, the conditions (\ref{con1}) and (\ref{con2}) suffice to satisfy the
condition (\ref{tobo}). This fact suggests that, on physical grounds, the term
$\int_{0}^{T}\left(  \partial_{r}j\right)  dv\ $should give the main
contribution to the total Bondi flux. It will be a self-consistency test of
these computations to confirm this expectation by showing that in the formula
of the final mass only the term $\int_{0}^{T}\left(  \partial_{r}j\right)  dv$
appears explicitly. Then, the problem reduces to solve Eqs. (\ref{F1}) and
(\ref{F2}) with the boundary conditions
\begin{equation}
m,j\in C^{2}\left(  \left]  0,T\right]  \times\left]  R^{\ast},\infty\right[
\right)  ,\label{derivabilita}%
\end{equation}%
\begin{equation}
\left|  m\right|  ,\left|  j\right|  \leq K_{1}<+\infty,\left|  \partial
_{v}j\right|  ,\left|  \partial_{r}m\right|  ,\left|  \partial_{r}j\right|
\underset{r\rightarrow+\infty}{\leq}K_{2}r^{-2}\label{ADM}%
\end{equation}%
\begin{equation}
\underset{r\rightarrow\infty}{\lim}\left|  \int_{0}^{T}\left(  \partial
_{r}j\right)  dv\right|  \leq K_{4}r^{-2},\mathtt{\ }\label{total}%
\end{equation}%
\[
m(0,r)=M>0;\,j(0,r)=k_{1};\,m(T,r)=\overline{M}\geq0;j(T,r)=k_{2}%
,\mathtt{\quad}%
\]
where $K_{1}$, $K_{2}$, $K_{4}$, $k_{1}$ and $k_{2}$ are positive constants.
It is important to stress that Eqs (\ref{F1}) and (\ref{F2}), not depending on
the explicit form of $\rho$ and $V$, are not coupled with the matter source
but, however, because of the relation (\ref{algebra}), are a direct
consequence of the non vanishing trace anomaly.

\section{The non vanishing final mass \ \ }

Now, it will be proven that the trace anomaly prevents a complete evaporation.
Indeed, by using the boundary conditions $j(0,r)=0$ and $m(0,r)=2M-\frac
{Q^{2}}{r}$, from Eq. (\ref{F1}) it follows that the function $m$ can be
expressed in closed form in terms of $j$ in the following way:
\begin{equation}
m(v,r)=r-\frac{r}{\frac{r^{2}}{r^{2}-2rM+Q^{2}}-\frac{1}{2}\partial_{r}%
\int_{0}^{v}\exp\left[  j(\tau,r)\right]  d\tau}\mathtt{{,} } \label{rapu}%
\end{equation}
whose time derivative reads
\begin{equation}
\partial_{v}m=-\frac{re^{j}\partial_{r}j}{2\left(  \frac{r^{2}}{r^{2}%
-2rM+Q^{2}}-\frac{1}{2}\partial_{r}\int_{0}^{v}\exp\left[  j(\tau,r)\right]
d\tau\right)  ^{2}}\mathtt{{.} } \label{decresce}%
\end{equation}
Thus, thanks to Eq. (\ref{stop}), the sign of $\partial_{v}m$\ is
directly related to the sign of $\rho$. The final mass
$\overline{M}$ may be expressed as
\begin{equation}
\overline{M}=\frac{1}{2}\left[ \frac{Q^{2}}{r}+r-\frac{r}{\frac{r^{2}}%
{r^{2}-2rM+Q^{2}}-\frac{1}{2}\partial_{r}\int_{0}^{T}\exp\left[
j(\tau,r)\right]  d\tau}\right] . \label{FM}%
\end{equation}

It is worth to note that $m(v,r)$, and then $\overline{M}$,\ cannot be
negative. In fact, by using Eqs (\ref{ADM}) and (\ref{FM}) we have:
\[
m\underset{r\rightarrow+\infty}{\sim}r-\frac{r}{1+2\frac{M}{r}-\frac{Q^{2}%
}{r^{2}}-\frac{vK_{2}\exp\left[  K_{1}\right]  }{2r^{2}}}\geq0
\]
being $M>0$. Furthermore, by Eq. (\ref{decresce}), it is trivial to show
that:
\begin{align*}
\rho &  \geq0\Rightarrow\partial_{r}j\leq0\Rightarrow\overline{M}\geq M\\
\rho &  <0\Rightarrow\partial_{r}j>0\Rightarrow\overline{M}<M\mathtt{.}%
\end{align*}
Then, when $\rho\geq0$, so that $T_{\mu\nu}$ describes a null scalar field
violating, at most, the \emph{weak} energy condition but not the \emph{null}
one, the evaporation does not take place. Moreover, in this case, the boundary
conditions (\ref{ADM}) do not play any role.

However, even when $\rho<0$, a remnant is left. In fact, Eq. (\ref{FM})
implies that the function
\[
F(r)=\frac{1}{2}\int_{0}^{T}\exp\left[  j(v,r)\right]  dv
\]
has to satisfy the following ordinary differential equation:
\begin{equation}
\partial_{r}F=\frac{r^{2}}{r^{2}-2rM+Q^{2}}-\frac{r^{2}}{r^{2}-2r\overline
{M}+Q^{2}}\mathtt{.} \label{L1}%
\end{equation}
Since the initial blackhole is not extremal, we can write
\begin{equation}
\frac{r^{2}}{r^{2}-2rM+Q^{2}}=\frac{1}{r_{+}-r_{-}}\left(  \frac{r^{2}%
}{r-r_{+}}-\frac{r^{2}}{r-r_{-}}\right)  \label{L2}%
\end{equation}
where
\[
r_{\pm}=M\pm\sqrt{M^{2}-Q^{2}}\mathtt{.}%
\]
Eqs (\ref{L1}) and (\ref{L2}) give:
\begin{equation}
F=r+\frac{1}{r_{+}-r_{-}}\ln\frac{\left(  r-r_{+}\right)  ^{r_{+}^{2}}%
}{\left(  r-r_{-}\right)  ^{r_{-}^{2}}}-\int^{r}\frac{x^{2}dx}{x^{2}%
-2x\overline{M}+Q^{2}}+N \label{zero}%
\end{equation}
where $N$ is an integration constant.

The above relation is not compatible with the boundary conditions. In fact,
unless $\overline{M}=M$, it is easy to see that
\[
F(r)\underset{r\rightarrow+\infty}{\rightarrow}\infty,
\]
while the boundary conditions (\ref{ADM}) imply
\[
\infty>\frac{1}{2}Te^{K_{1}}\geq\frac{1}{2}\int_{0}^{T}e^{j(v,r)}%
dv=F(r)\underset{r\rightarrow+\infty}{\rightarrow}\infty\mathtt{.}%
\]
Furthermore, main qualitative features do not change if we also allow a
variation of the charge, the result being always incompatible with the
formation of a naked singularity. Then, even with the ''worst''
energy-momentum tensor (\emph{i.e.} a $T_{\mu\nu}$ violating all the energy
conditions), the surface gravity cannot vanish preventing the formation of a
naked singularity, that is, the evaporation stops because of the trace of the
energy-momentum tensor that carries away energy from the blackhole to infinity
\cite{CV303}. Moreover, thanks to the boundary condition (\ref{total})\emph{,}
it is easy to see that the same results also hold in the limit $T\rightarrow
\infty$. In fact, Eq. (\ref{FM}) shows that, in order to have a final mass
different from the initial one, the following relation should hold:
\begin{equation}
\int_{0}^{\infty}\exp\left[  j(v,r)\right]  \partial_{r}jdv\underset
{r\rightarrow+\infty}{\sim}\frac{K_{5}}{r},\label{forse}%
\end{equation}
$K_{5}$ being a positive constant (otherwise, the coefficient of
the $1/r$-term in the denominator of the r.h.s. of Eq. (\ref{FM}),
that represents the blackhole mass, would not change). However,
from relation (\ref{total}), we get
\[
\int_{0}^{\infty}\exp\left[  j(v,r)\right]  \partial_{r}jdv\leq e^{K_{1}}%
\frac{K_{4}}{r^{2}},
\]
and the above relation is clearly incompatible with Eq. (\ref{forse}). By
taking the limit of a vanishing charge \cite{CV303}, we obtain that the trace
anomaly of the Hawking particles is able to stop the evaporation
irrespectively of the positivity conditions fulfilled by the energy-momentum tensor.

In conclusion, in this toy model, it could be possible to take
care of the puzzles raised by the Hawking radiation, such as the
possible violations of the third law, formation of naked
singularities, the \emph{unitarity puzzle} and the
\emph{information} \emph{loss} \emph{paradox} \cite{CV303}, by
taking into account the trace anomaly \emph{via} the exact
Einstein equations. This is a quite interesting results since the
trace anomaly, that is a generic feature of QFT (namely, it is
related to the $\beta-$function of the quantum fields that can
vanish only in highly idealized (super)symmetric theories), has a
prominent role in stopping the evaporation, so, in some sense, QFT
itself could prevent the formation of a naked singularity and the
loss of unitarity. This would be a positive self-consistency check
for QFT \cite{CV303}. To complete the analysis, the explicit
expression, in terms of the charge $Q=Ze$, of the remnant mass
$\overline{M}=M_{P}f\left(  Z\right)  $ should be determined,
where $M_{P}=\sqrt{\hbar c/G} $ denotes the Planck mass. At this
level of analysis, in which numeric computations have not yet been
performed, we may only say that $f\left(  0\right)  \neq0$, since
previous results also hold in the limit $Q\longrightarrow0$. The
function $f$ strongly depends on the conformal anomaly whose
expression, being determined by all quantum fields that contribute
to the Hawking radiation, is not available yet. However, the
remnant mass is an increasing function of conformal anomaly
because this latter is responsible for stopping the evaporation.

Many previous works on this topic \cite{Ba81}, \cite{Ma95} showed that, by
neglecting the four-dimensional trace anomaly, the blackhole, in a time of the
order $O(M^{3})$, can loose all its mass, as first predicted by Hawking
\cite{Ha74}. Thus, the following questions arises:

\emph{which is the new insight related to the four dimensional trace anomaly?}

\emph{when is the trace anomaly not negligible anymore?}

\section{Field-theoretical features}

In this section the Planck constant $\hbar$, the light velocity
$c$ and the gravitational constant $G$ have been restored. It is
easy to check that Eqs (\ref{F1}) and (\ref{F2}) are formally
scale-invariant. Namely, if $m_{0}$ and $j_{0}$ are a solution of
the system, then $\lambda m_{0}$ and $j_{0}$ are a solution too
(provided we also make the transformation $r\rightarrow\lambda r$,
$t\rightarrow\lambda t) $. Then, because of the freedom in
changing the energy scale, Eqs (\ref{F1}) and (\ref{F2}) by
themselves do not provide us with a lower bound on the remnant
mass.

However, this is only formally true. For instance, the presence of
a nonvanishing cosmological constant $\Lambda$, being the value of
$\Lambda$ fixed by experiments, would violate the scale invariance
because no scale transformation on $\Lambda$\ could be performed.
In other words, $\Lambda$ introduces a characteristic energy scale
in the evaporation process. The same is true for the trace
anomaly, because it also breaks the scale invariance and this
breaking is hidden in the r.h.s. of the Einstein equations.

Now we will present a highly qualitative argument to obtain a lower bound for
the remnant mass and a new insight into the relations between a complete
evaporation and the loss of unitarity on purely thermodynamical grounds,
although, in order to know exactly when the trace anomaly comes into play, one
should have the explicit expressions of both $\rho$ and $V$ or, at least, an
estimate of their ratio $V/\rho$. Furthermore, in order to provide a clear
physical intuition, we will follow as close as possible an interesting analogy
with the computation of the Casimir force. In the following, for the sake of
simplicity, we will consider the case of a Schwarzschild blackhole shortly
analysed in \cite{CV303}, although the analysis can be extended to the charged
case too. The treatment is absolutely non rigorous since it is intended only
to give an intuitive picture of the process.

As originally found by Hawking \cite{Ha74}, the expectation value of the
operator number of a bosonic field of spin zero, measured by a static observer
in the asymptotic future of Schwarzschild blackhole of mass $M$, is
\begin{equation}
n_{i}=\frac{1}{\exp(\hbar\omega_{i}/kT_{H})-1},\label{radhaw}%
\end{equation}
where $k$ is the Boltzmann constant and
\[
T_{H}=\frac{\hbar c^{3}}{8\pi kGM}%
\]
is the Hawking temperature. Since our purpose is to get only an order of
magnitude estimate, in the following the \emph{greybody} \emph{factors} will
be neglected.

It is known that the basic ingredients to get the relation (\ref{radhaw}) are
the existence of a horizon and the fact that on curved space-times, because of
the lacking of a preferred Poincar\'{e} symmetry group, the notion of particle
is not an invariant concept. Thus, since the Einstein equations are not needed
to get the above relation, it is lawful to say that the Hawking effect is only
''kinematical'' \cite{Vi98}. From the physical point of view, this means that,
since in deducing relation (\ref{radhaw}) the background metric is kept fixed,
we are neglecting the backreaction effects of the emitted particles on the
metric so that the energy conservation is not fulfilled. In fact, the
blackhole should give to the Hawking particles the energy they need to be
created but, if $M$ does not change, this, obviously, does not happen.

Hence, by imposing the conservation of energy, we will take into
account the backreaction, \emph{i.e.} we will ensure that this
thermal particles are created by the mass lost by the blackhole.
In the following, we will assume that the expectation value of the
operator number $n_{i}=n_{i}\left( M\right)  $ changes
adiabatically as a function of $M$. This assumption is well
verified in all the situations in which the blackhole
thermodynamics is applicable \cite{Ma95}.

Now we will compute how the expectation number (\ref{radhaw}) increases due to
the decrease of the blackhole mass and then we will write the conservation of
energy for this process. This point of view is quite similar to the one
adopted in the computation of the \emph{Casimir force}. In fact, in the latter
case, one first computes the zero modes and the vacuum energy (that is, of
course, a divergent but unobservable quantity) in a configuration with some
nontrivial boundaries, and then computes the variation of the vacuum energy
due to an infinitesimal change in the boundaries that is finite and observable
as well. In the blackhole case, the boundary is the Schwarzschild radius so
that a change in the mass can be thought as a change in the boundary. For this
reason, we should do the same and obtain the same results (namely, that the
energy due to the particles created by the mass lost by the blackhole is a
finite quantity). Therefore, if $M\rightarrow M+\delta M$ , then
$n_{i}\rightarrow n_{i}+\delta n_{i}$ with
\begin{equation}
\delta n_{i}=\frac{\partial n_{i}}{\partial M}\delta M=-\frac{8\pi
G\hbar\omega_{i}\exp(\hbar\omega_{i}/kT_{H})}{\hbar c^{3}\left(  \exp
(\hbar\omega_{i}/kT_{H})-1\right)  ^{2}}\delta M\mathtt{.} \label{delta}%
\end{equation}
Then, the main step to get a lower bound of the remnant mass is to impose that
the new particles are created by the mass lost by the black-hole, that is, we
have to fulfil the conservation of energy:
\begin{equation}
c^{2}\delta M+\sum_{i}\hbar\omega_{i}\delta n_{i}\leq0, \label{rad1}%
\end{equation}
where, since the blackhole is evaporating, $\delta M<0$\ and in Eq.
(\ref{rad1}) the inequality has to be used. In fact, besides the spin-zero
bosons described by the relation (\ref{radhaw}), the mass lost by the
blackhole also gives rise to other kind of particles, such as fermions or
bosons with higher spin. For this reason, if we would use the equality in Eq.
(\ref{rad1}), we would neglect all the other particles that the blackhole is
able to create. Then, the second term in Eq. (\ref{rad1}) encodes, in the
thermodynamical limit, the backreaction effects of the emitted particles on
the gravitational field. It is clear from Eq. (\ref{rad1}) that, as long as
$M$ is big enough, the second term in the inequality is completely negligible
and the usual blackhole thermodynamics should apply.

However, the smaller is $M$, the harder is to fulfil Eq. (\ref{rad1}). In
fact, from Eqs. (\ref{delta}) and Eq. (\ref{rad1}) it follows that
\begin{equation}
1\geq\frac{8\pi}{\left(  c^{2}M_{P}\right)  ^{2}}\sum_{i}\frac{\left(
\hbar\omega_{i}\right)  ^{2}\exp(\hbar\omega_{i}/kT_{H})}{\left(  \exp
(\hbar\omega_{i}/kT_{H})-1\right)  ^{2}}, \label{rad2}\mathtt{{.} }%
\end{equation}

It is worth to note here that, in the above relation, as expected on the
grounds of the analogy with the Casimir case, neither infrared nor ultraviolet
divergences appear, so that we do not need any regularization. Furthermore,
Eq. (\ref{rad2}) implies that the conservation of energy can be satisfied only
for $M\geq\overline{M}$, where $\overline{M}$ is implicitly defined by the
following equation:
\[
1=\frac{8\pi}{\left(  c^{2}M_{P}\right)  ^{2}}\sum_{i}\frac{\left(
\hbar\omega_{i}\right)  ^{2}\exp(8\pi G\omega_{i}\overline{M}/c^{3})}{\left(
\exp(8\pi G\omega_{i}\overline{M}/c^{3})-1\right)  ^{2}},
\]
It is worth noting here that this is only a lower bound and that
the remnant could be quite larger than the Planck mass. In fact,
if we would be able to include in Eq. (\ref{rad1}) all the
different kinds of massless particles that contribute to the
evaporation process (such as fermions or bosons with higher spin,
etc.), then the inequality in Eq. (\ref{rad1}) would become an
equality, so that there would not be phase space available anymore
for the evaporation itself. Moreover, as it has been shown above,
if the trace anomaly $V$ was of the same order of $\rho$ then the
evaporation would halt well before reaching the Planck scale.

Thus, the conservation of energy, and then the backreaction,
prevents the blackhole mass $M$ from being lower than
$\overline{M}$. Since a smaller mass is not allowed by the energy
conservation, the remnant is stable. Besides being an interesting
result in itself, this makes clear the relation between energy
conservation and complete evaporation. In fact, it has been
speculated \cite{Ha76} that, since it seems that a complete
evaporation cannot be ruled out, QFT has to be generalized to
allow nonunitary processes. However, if one tries to do this, then
the energy conservation is lost \cite{El84}, \cite{Ba84}. Instead,
Eq. (\ref{rad2}) shows that the energy conservation itself stops
the evaporation. Moreover, this conclusion fits very well with the
results obtained in \cite{Ca01}, \cite{Ah87}.

\section{ Conclusions and perspectives}

In this paper a toy model is proposed to describe a transition
from a Reissner-Nordstrom blackhole of mass $M$ and charge $Q$ to
a Reissner-Nordstrom blackhole of mass $\overline{M}$ and charge
$Q$. The basic equations are derived from the non-vacuum Einstein
equations, the source being an energy-momentum tensor with a non
vanishing trace representing the massless quanta of the Hawking
particles. Nevertheless, the results do not depend on the explicit
form of the energy-momentum tensor, thus allowing in principle
arbitrary violations of all the \emph{energy conditions}. We
showed, \emph{via} the exact Einstein equations, that the trace of
the energy-momentum tensor prevents any lowering of the surface
gravity saving, in this way, the third law of blackhole
thermodynamics from the Hawking radiation and preventing the
formation of a naked singularity. In the limit of a vanishing
charge, this approach shows that the trace anomaly of the Hawking
particles stops the evaporation. In this way, the
\emph{information} \emph{loss} \emph{paradox} could be resolved by
taking into account the information detained by the remnant and by
the Hawking radiation through the Bekenstein mechanism
\cite{Be93}. We also briefly analyzed the field-theoretical
connection between the blackhole evaporation and the cosmological
constant.

\section*{Acknowledgments}

The authors wish to thank D.Grumiller, G.Esposito and S.Sonego for remarks and
important bibliographic suggestions.

\end{document}